\documentclass[sigconf]{acmart}

\usepackage{algorithmic}
\usepackage{graphicx}
\usepackage{textcomp}
\usepackage{xcolor}
\def\BibTeX{{\rm B\kern-.05em{\sc i\kern-.025em b}\kern-.08em
    T\kern-.1667em\lower.7ex\hbox{E}\kern-.125emX}}

\usepackage{url}

\usepackage{booktabs}

\usepackage{cleveref}[2012/02/15]
\crefformat{footnote}{#2\footnotemark[#1]#3}
\usepackage[linesnumbered,commentsnumbered,ruled,vlined]{algorithm2e}

\usepackage{color,soul}

\usepackage{ulem}
\usepackage{makecell}
\usepackage{tcolorbox}

\begin{document}

\title{Contrastive Learning-Enhanced Large Language Models for Monolith-to-Microservice Decomposition}

\author{Khaled Sellami}
\email{khaled.sellami.1@ulaval.ca}
\orcid{0000-0002-6595-2489}
\affiliation{%
  \institution{Laval University}
  \city{Québec}
  \state{QC}
  \country{Canada}
}

\author{Mohamed Aymen Saied}
\email{mohamed-aymen.saied@ift.ulaval.ca}
\orcid{0000-0002-9488-645X}
\affiliation{%
  \institution{Laval University}
  \city{Québec}
  \state{QC}
  \country{Canada}
}

\begin{abstract}
  As Monolithic applications evolve, they become increasingly difficult to maintain and improve, leading to scaling and organizational issues. The Microservices architecture, known for its modularity, flexibility and scalability, offers a solution for large-scale applications allowing them to adapt and meet the demand on an ever increasing user base. Despite its advantages, migrating from a monolithic to a microservices architecture is often costly and complex, with the decomposition step being a significant challenge. This research addresses this issue by introducing MonoEmbed, a Language Model based approach for automating the decomposition process. MonoEmbed leverages state-of-the-art Large Language Models (LLMs) and representation learning techniques to generate representation vectors for monolithic components, which are then clustered to form microservices. By evaluating various pre-trained models and applying fine-tuning techniques such as Contrastive Learning and Low Rank Adaptation (LoRA), MonoEmbed aims to optimize these representations for microservice partitioning. The evaluation of the fine-tuned models showcases that they were able to significantly improve the quality of the representation vectors when compared with  pre-trained models and traditional representations. The proposed approach was benchmarked against existing decomposition methods, demonstrating superior performance in generating cohesive and balanced microservices for monolithic applications with varying scales. 
\end{abstract}

\settopmatter{printacmref=false}
\renewcommand\footnotetextcopyrightpermission[1]{}
\settopmatter{printacmref=false, printccs=false, printfolios=false}
\acmConference[ ]{ }{ }{ }
\acmBooktitle{ }
\acmPrice{}
\acmISBN{}
\acmDOI{}
\fancyhead[LE]{}   % Left side of even pages
\fancyhead[RO]{}   % Right side of odd pages


\maketitle

% \section{Introduction}
\section{Introduction}
A Monolithic Architecture refers to software architectures where all of the application’s components are deployed as a single unit \cite{newman2019monolith}. As the scale of the monolithic applications gets larger, maintaining such software becomes extremely complex while scaling and organizational issues start emerging \cite{soundcloud2024}. Due to a multitude of factors such as its modularity and its compatibility with Cloud technologies, the Microservices architectural style has proven to be effective in meeting the needs of applications deployed at large scales \cite{spotify2022,netflix2015,soundcloud2024,vayghan2019availability,vayghan2021stateful,almarimi2019apirec}. However, migrating from an established monolithic system to a more modern and more complex microservices architecture has been shown to be an extremely expensive and difficult process \cite{rosati2019migrationcost,fritzsch_2019_migrationindustry,kalske2018m2mchallenges,francesco_2018_idustrialsurvey,vayghan2018experience}. In-depth analysis of the migration pipeline \cite{faustino2022stepwise,mazzara2021casestudy} has identified the decomposition step, which consists of partitioning the monolith's component into target microservices, as the largest and most important roadblock when refactoring a monolith into microservices. 

There has been an increasing interest \cite{Abgaz2023decompsurvey,oumoussa2024decompsurvey} in automating the decomposition process to reduce its constraints and to democratize it. Decomposition approaches range from semi-automated methods \cite{gysel2016servicecutter,li2019dataflowdec} which aim to guide the developers through the process to mostly automated approaches \cite{mathai2022chgnn,jin2021fosci,kalia2021mono2micro,khaled2022hydecomp,yedida2023deeply,aldebagy2021code2vec} that generate decompositions using input representations of the monolith such as its source code or its design artifacts. 

In fact, one of the most common approaches to analyzing the monolithic applications for the task of decomposing them to microservices is the static analysis of the source code \cite{Abgaz2023decompsurvey}. However, this analysis method has often under-performed when compared with dynamic analysis for example \cite{desai2021cogcn,kalia2021mono2micro}. On the other hand, these alternative representations of the monolith often come with their own disadvantages such as low coverage in the case of dynamic analysis. While there has been some effort into combining multiple representations \cite{mathai2022chgnn,khaled2022hydecomp,qian2023gdcdvf}, most analysis techniques in microservices decomposition research has relied mainly on traditional methods (e.g. static and dynamic analysis) and explicitly defined features (e.g. cohesion, coupling). On the other hand, recent software engineering research has showcased the potential of using Large Language Models (LLMs) \cite{niu2023codeembedreview} and Representation Learning techniques \cite{guo2022unixcoder,wang2023codet5p} to enhance the performance of downstream tasks \cite{Torregrossa2023embeddingssurvey} like code summarization and code search \cite{junkai2024codesearch} through rich, high-dimensional representation vectors of source code.

In this research, we present MonoEmbed, a Language Model based approach for decomposing monolithic applications into microservices. MonoEmbed consists of a pipeline with two main components: Analysis and Inference. The Analysis component contains the Language Model and is responsible for transforming the monolith's elements, in the form of source code fragments, into efficient embedding vectors that were optimized for the decomposition problem. We reviewed a large and diverse number of Pre-Trained Language Models (e.g. Encoders \cite{guo2021graphcodebert} , LLMs\cite{aimeta2024llama3modelcard}) and improved their performance through a Contrastive Learning based fine-tune process (i.e. by maximizing the distance between classes of different microservices while reducing the differences between classes that should be in the same microservice). By imporving the quality of these embedding vectors, we enable better microservices suggestions in the Inference component which utilizes clustering algorithms to partition the classes.

Using a large set of existing microservices applications, we train the Language Models and evaluate the quality of the embeddings they generate based on their similarity with the true partitions. We compare the performance of a large number of clustering algorithms with different models. We evaluate the performance of MonoEmbed in comparison with decomposition benchmark approaches on 8 monolithic applications. Results showcase that our approach is able to generate more cohesive and consistent decompositions. 

The main contributions of this research are:
\begin{itemize}
    \item We propose a Language Models based decomposition approach and show that it generates more consistent and cohesive results on applications with varying scales. 
    \item We evaluate 18 Pre-Trained Language Models and 4 benchmark representations to review the quality of their embedding vectors in the decomposition space.
    \item We introduce and evaluate potential fine-tuning methods and datasets for decompositions. Our Fine-Tuned LLMs outperform current models, producing robust embeddings that effectively distinguish microservice boundaries.
\end{itemize}

% \section{Related Work}

% \subsection{Motivating example}
\section{Motivating Example}\label{subsec:mexample}

A decomposition approach survey \cite{Abgaz2023decompsurvey} reveals that most monolithic application decomposition methods rely on static code analysis to extract component relationships, primarily between classes. Common approaches include generating Abstract Syntax Trees (AST) and call graphs, where nodes represent classes and edges represent invocations \cite{khaled2022hydecomp,desai2021cogcn}. The resulting adjacency matrix, referred to as \textit{ST-Calls}, represents the monolith's features. Semantic representations using Bag-of-Words (\textit{SM-BoW}) or Term Frequency - Inverse Document Frequency (\textit{SM-TFIDF}) vectors are commonly used as well.

\begin{figure*}
\centering
\includegraphics[width=0.9\linewidth]{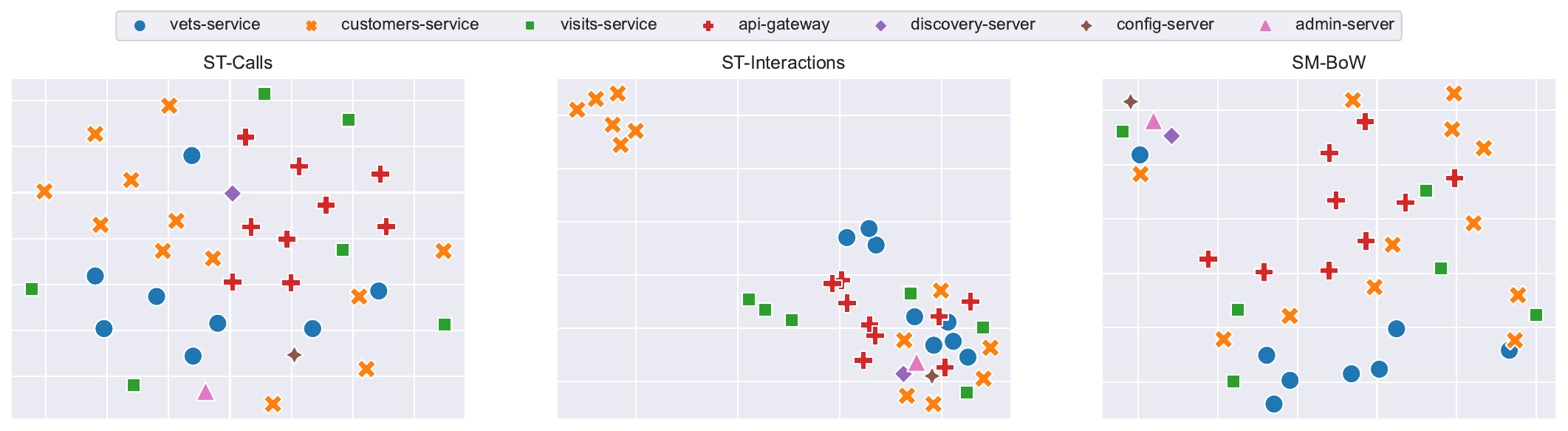}
\caption{A UMAP 2-dimensional projection of calls, interactions and bag-of-words representations.}
\label{fig:mexample}
\end{figure*}

Fig \ref{fig:mexample} illustrates a 2-D projection of three representations for the Spring PetClinic Microservices application \cite{microapps2024petclinic}, a microservices variant of a benchmark in decomposition research \cite{jin2021fosci,khaled2022hydecomp}. The figure shows classes as points, with colors and shapes indicating their respective microservices. The representations include ST-Calls, ST-Interaction (combining calls, inheritance, and variable references) \cite{khaled2022hierdecomp,khaled2022hydecomp}, and SM-BoW. The projection was generated using the Uniform Manifold Approximation and Projection (UMAP) algorithm \cite{mcinnes2020umapuniformmanifoldapproximation}. We can see that across all representations, most classes are scattered, making it difficult to discern clear microservice groupings. However, the \textit{ST-Interactions} representation captures relationships between some \textit{Customers} service classes. While the \textit{SM-BoW} representation shows groupings of some \textit{Vets}, \textit{Customers} and \textit{API-Gateway} microservice classes, they are too close to each other to differentiate clearly.
On the other hand, some groupings don't reflect actual microservices, as seen in the top-left cluster of \textit{SM-BoW}. Here, seven classes from different microservices cluster together. Upon closer inspection, these classes are service entry points with nearly identical code, differing only in names. It highlights a weakness of this semantic representation approach. All these representations have weaknesses that could hinder decomposition approaches. Their simple encodings alone are insufficient to create domain-relevant representations. Which is why we propose an approach based on Language Model representations.

\section{Background and Related Work}

% \subsection{Background}
\subsection{Background}
\subsubsection{Embedding Models (EMs)} and in particular Neural-Network (NN) based Embedding Models like Word2Vec \cite{mikolov2013word2vec1} transform input sequences (e.g. word tokens, code samples) into continuous vector representations in a high-dimensional space. These embedding vectors capture semantic and syntactic relationships between the tokens, allowing similar concepts to have similar vector representations. 

\subsubsection{Language Models (LMs)} are probabilistic models that learn to predict the likelihood of a sequence of words. Modern LMs are based on NN architectures like Recurrent Neural Networks (RNN) \cite{alon2019code2seq} and Transformers \cite{devlin2019bert,liu2019roberta,radford2018gpt1}. Large Language Models (LLMs) often refers to Pre-Trained Neural LMs based on the Transformers \cite{vaswani2017attention} architecture and have more than 1 Billion parameters. While LMs and EMs are not the same, recent NN-based LMs such as BERT \cite{devlin2019bert} and GPT \cite{radford2018gpt1} can generate rich contextual representations and, therefore, can be utilized as EMs. For the rest of the paper, we will use both terms interchangeably.

\subsubsection{Transformers} are advanced NN architectures designed to handle sequential data using the self-attention mechanism \cite{vaswani2017attention}.  Transformers consist of an encoder-decoder structure, where the encoder processes the input sequence to generate contextual embeddings and the decoder uses these embeddings to produce the output sequence. While encoder-decoder transformers \cite{raffel2020t5} are versatile for a wide range of sequence-to-sequence tasks, encoder-only transformers, such as BERT \cite{devlin2019bert}, are more often used as EMs to generate efficient contextual embeddings. Decoder-only Transformers, such as the GPTs \cite{radford2018gpt1,brown2020gpt3}, have proven to be most successful on generative problems. However, recent efforts have been able to adapt the LLM decoder-only transformers into EMs \cite{springer2024llmrepetition,parishad2024llm2vec,muennighoff2024gritlm,SFRAIResearch2024sfrmistral} in order to take advantage of their capabilities.

\subsubsection{Contrastive Learning (CL)} is a self-supervised representation learning approach where an EM is trained to recognize similarities and differences between data samples. The core idea is to push similar or related samples (positive pairs) closer together in the embedding space, while pushing dissimilar samples (negative pairs) farther apart.

\subsection{Related Work}
Most decomposition approaches rely on static analysis or dynamic analysis in order to extract microservices from a monolithic application. For example, TopicDec \cite{brito2021topicmodeling} relies on Topic Modeling methods and static analysis to extract domain and syntax level representation of the classes in a monolith. In a similar fashion, HierDec \cite{khaled2022hierdecomp} and MSExtractor \cite{khaled2022msextractor,saidani2019msextractor} apply static analysis and traditional Natural Language Processing pipelines to extract the structural and semantic relationships between the classes. On the other hand, approaches such as FoSCI \cite{jin2021fosci} and Mono2Micro \cite{kalia2021mono2micro} rely mainly on dynamic analysis in conjunction with clustering or genetic algorithms to recommend microservices. HyDec \cite{khaled2022hydecomp} attempts to combine both analysis methods and avoid their drawbacks. Other methods have used different inputs and representations such as MEM \cite{mazlami2017mem} which relied on the version history and software evolution \cite{benomar2015evolution} and the semi-automated methods \cite{daoud2023multimodeldec,li2019dataflowdec} like ServiceCutter \cite{gysel2016servicecutter} which utilized the design artifacts or dataflow diagrams. More recently, there has been an interest in incorporating the databases in the decomposition as seen by the approaches CHGNN \cite{mathai2022chgnn}, CARGO \cite{vikram2022cargo} and DataCentric \cite{yamina2022datacentric}.

Several Neural Network-based approaches have been proposed for the decomposition problem. CoGCN \cite{desai2021cogcn} employs a Graph Neural Network (GNN) to learn class partitioning and detect outliers through dynamic analysis. Deeply \cite{yedida2023deeply} and CHGNN \cite{mathai2022chgnn} enhance this method with hyper-parameter tuning, a novel loss function and additional input representations while GDC-DVF \cite{qian2023gdcdvf} combines structural and business representations in its GNN-based approach. MicroMiner \cite{trabelsi2023microminer} offers a 3-step framework, utilizing CodeBERT \cite{feng2020codebert} embeddings and a classifier to categorize source code fragments. In contrast, Code2VecDec \cite{aldebagy2021code2vec} leverages the Code2Vec \cite{alon2018code2vec} embedding model to generate class vectors, which are then used for clustering. While GNN-based decomposition methods incorporate some representation learning, it's limited to their training applications and input analysis approaches. For example, CoGCN's encoder learns class embeddings but requires training for each application based on structural and dynamic analysis results. To our knowledge, only MicroMiner and Code2VecDec have utilized Transformers or static embedding models in decompositions. Our approach extends this by comparing the performance of various models, including state-of-the-art Large Language Models, and proposing a fine-tuning method to enhance their effectiveness. This strategy aims to deepen the use of such models in addressing the decomposition problem, going beyond the limitations of existing methods.

\section{Methodology}\label{sec:approach}

\subsection{Overview}\label{subsec:overview}
\begin{figure}[ht]
\centering
\includegraphics[width=\linewidth]{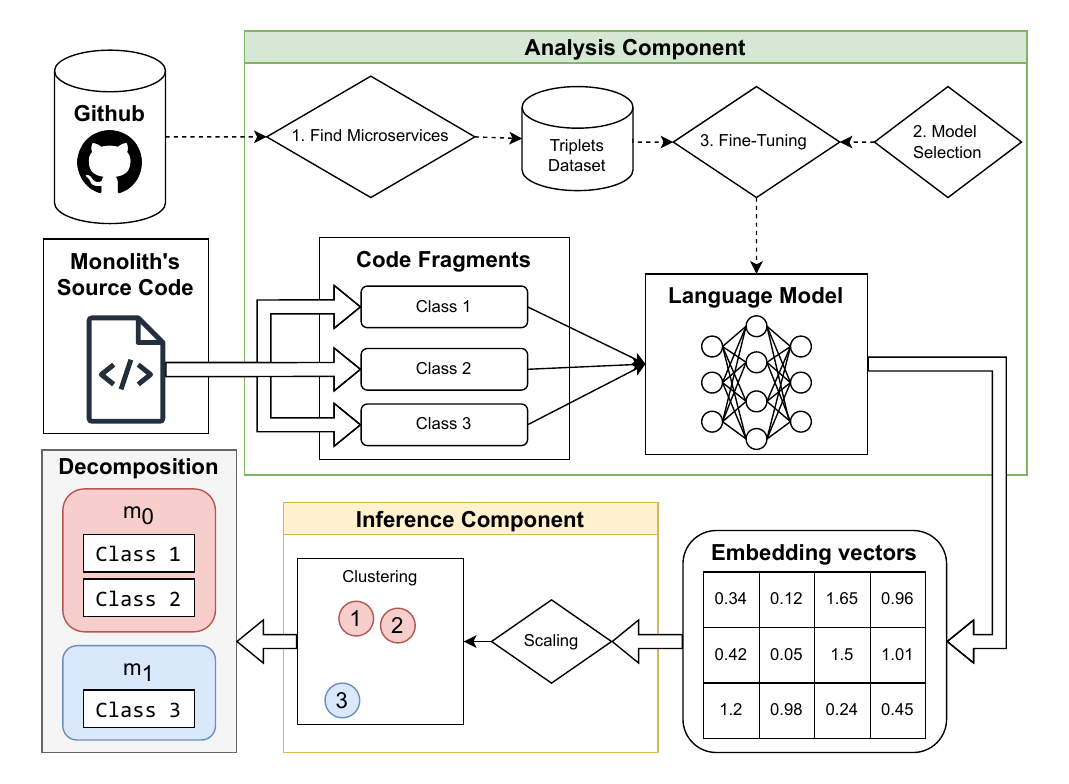}
\caption{An overview of the suggested decomposition approach.} \label{fig:overview}
\end{figure}

MonoEmbed follows, overall, the same structure as most decomposition approaches. Fig     \ref{fig:overview} showcases two main components: \textbf{Analysis} and \textbf{Inference}. The design and application of our approach is done in two phases.

The first phase revolves around training and preparing the LM of the \textbf{Analysis} component. It involves the creation of a triplet samples dataset based on Microservices applications, the selection between different Pre-Trained LMs and the fine-tuning of the model through a Contrastive Learning method.

The second phase showcases the usage of MonoEmbed on monolithic applications with the Fine-Tuned LM. In this phase, the \textbf{Analysis} component processes the monolith's source code using the LM, generating a feature matrix  \( (N{\times}M) \), where \( (N) \) is the number of classes and \( (M) \) is the embedding vector size. The \textbf{Inference} component then normalizes and partitions this data with a clustering algorithm to produce the decomposition, a set of microservices with distinct classes. For this research, we limit our input to source code at the class level granularity.

% \subsection{Selecting the Embedding Models}
\subsection{Model Selection}\label{subsec:mselection}

While various analysis methods could satisfy our definition of the analysis component, even including static analysis approaches, this research primarily focuses on utilizing Language Models to transform source code into meaningful feature vectors. A study of the usage of Pre-Trained Models \cite{niu2023codeembedreview} in Software Enginnering in 2023 showcases the large number of potential models, which has been growing significantly ever since. For this reason, the selection of an appropriate LM from the diverse array of options is complex, considering factors such as source code nature, desired representation granularity, and computational efficiency, all of which can impact decomposition task performance. To facilitate understanding and inform the selection process, we provide an overview of potential embedding models, categorizing them by types and characteristics. Table \ref{tab:ptmodels} summarizes the models used and their respective categories. It includes the following columns: \textit{Size} (approximate number of parameters), \textit{Context} (maximum input tokens), \textit{Code} (Pre-Trained on source code), and \textit{Modality} (acceptable input types for each model).

\begin{table}[]
\caption{Metadata of the potential Pre-Trained Language Models}
\label{tab:ptmodels}
\resizebox{\linewidth}{!}{%
\begin{tabular}{llllll}
\hline
Model Name                                              & Group           & Size & Context & Code & Modality    \\ \hline
Code2Vec \cite{alon2018code2vec}               & Static       & \textless{}1M & -    & Yes & AST paths \\
Code2Seq \cite{alon2019code2seq}               & RNN    & \textless{}1M & -    & Yes & AST paths \\
CuBERT \cite{kanade2020cubert}         & ET    & 340M & 2048        & Yes      & PL          \\
CodeBERT \cite{feng2020codebert}       & ET    & 125M & 512         & Yes      & NL-PL       \\
GraphCodeBERT \cite{guo2021graphcodebert}      & ET & 125M                           & 512  & Yes & DF-NL-PL  \\
UniXcoder \cite{guo2022unixcoder}              & ET & 125M                           & 1024 & Yes & DF-NL-PL  \\
CodeT5+ \cite{wang2023codet5p}         & EDT & 110M & 512         & Yes      & PL          \\
Meta Llama 3 \cite{aimeta2024llama3modelcard}     & DT & 8B                           & 8K  & No & NL        \\
SFR Mistral \cite{SFRAIResearch2024sfrmistral} & DT & 7B                             & 4096 & No  & NL        \\
E5 Mistral \cite{wang2024e5mistral}    & DT    & 7B   & 4096        & No       & NL          \\
CodeLlama \cite{rozière2024codellama}  & DT    & 7B   & 100K        & Yes      & NL          \\
DeepSeek Coder \cite{guo2024deepseekcoder}     & DT & 6.7B                           & 16K  & Yes & NL        \\
GritLM \cite{muennighoff2024gritlm}    & LME    & 7B   & 4096        & No       & Instruct-NL          \\
NVEmbed \cite{lee2024nvembed}          & LME             & 7B   & 4096        & No       & Instruct-NL \\
LLM2Vec \cite{parishad2024llm2vec}     & LME             & 8B   & 512         & No       & Instruct-NL \\
VoyageAI \cite{voyageai2024embeddings} & CEM   & -    & 16K         & No       & NL          \\
OpenAI \cite{openai2024embeddings}     & CEM   & -    & 8191        & No       & NL          \\
Cohere \cite{cohere2024embeddings}     & CEM   & -    & 512         & No       & NL          \\ \hline
\end{tabular}%
}
\end{table}

\subsubsection{Static and RNN based EMs}
\textit{Code2Vec} \cite{alon2018code2vec} is a static EM that was tailored for encoding code sample ASTs into static vector representations. It was extended by \textit{Code2Seq} \cite{alon2019code2seq}, a RNN-based EM, to incorporate input context.

\subsubsection{Encoder-only Transformers (ET)}
In our research, we considered several prominent ET models trained on software engineering problems and source code \cite{niu2023codeembedreview}. We selected \textit{CuBERT} \cite{kanade2020cubert} for its Java source code pairs modality. \textit{CodeBERT} \cite{feng2020codebert} and \textit{GraphCodeBERT} \cite{guo2021graphcodebert} were chosen for their unique multi-modal approach, incorporating Natural Language (NL), Programming Language (PL), and Data-Flow (DF) information. Additionally, we include \textit{UniXcoder} \cite{guo2022unixcoder}, which, although based on \textit{RoBERTa} \cite{liu2019roberta} like its predecessors, stands out as a unified model trained on both representation and generative tasks, offering enhanced potential for fine-tuning.
\subsubsection{Encoder-Decoder Transformers (EDT)}
\textit{CodeT5+} \cite{wang2023codet5p} is an Encoder-Decoder LLM that was trained on a large and diverse set of objectives so that it efficiently adapts to downstream tasks. We include its encoder in our evaluation.
\subsubsection{Decoder-only Transformers (DT)}
While DT LLMs are primarily designed for generative tasks, certain models \cite{wang2024e5mistral} have demonstrated strong performance in representation benchmarks, notably the Massive Text Embedding Benchmark (MTEB)\footnote{\label{footnote:mteb}\url{https://huggingface.co/spaces/mteb/leaderboard}} \cite{muennighoff2022mteb}. Our study encompasses a diverse range of DT models, as detailed in Table \ref{tab:ptmodels}, spanning from general-purpose models \cite{aimeta2024llama3modelcard} to those specialized for code-related tasks \cite{rozière2024codellama}.
\subsubsection{Language Model Embeddings (LME)}
We define LMEs as the DT LLMS that have been adapted for representation learning. We include state-of-the-art LMEs such as \textit{LLM2Vec} \cite{parishad2024llm2vec}, \textit{NV-Embed} \cite{lee2024nvembed} and \textit{GritLM} \cite{muennighoff2024gritlm}.
\subsubsection{Closed-source Embedding Models (CEM)}
CEMs refer to LLM based proprietary tools, like \textit{VoyageAI} \cite{voyageai2024embeddings}, designed to generate vector representations of text. They are accessed through APIs, enabling embedding generation without direct access to the underlying model. 

% \subsection{Fine-Tuning the Model}

\subsection{Fine-Tuning the Model}\label{subsec:finetune}

\begin{figure}
\centering
\includegraphics[width=\linewidth]{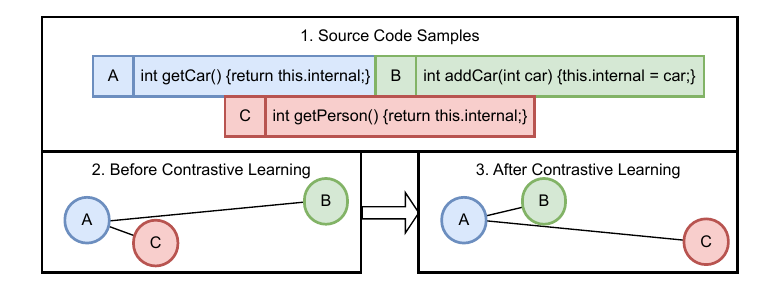}
\caption{An example of Contrastive Learning training.} \label{fig:clexample}
\end{figure}

The EM in our analysis is used to encode source code into vectors where classes from the same microservices have similar vectors, and those from different microservices are farther apart. Through the model selection step, we created a benchmark of Pre-Trained Models. Now, we propose a Contrastive Learning based fine-tuning approach to adapt these models and create embeddings suited for the decomposition task. Let's consider the example in Fig. \ref{fig:clexample}. Code samples \textit{A} and \textit{B} belong to the same microservice \textit{Car} while \textit{C} belongs to the microservice \textit{Person}. Due to the near-identical source code of \textit{A} and \textit{C}, Pre-Trained Model embeddings are closer together, with \textit{B} farther apart (Fig. \ref{fig:clexample} part 2). Using CL with a tailored objective function and dataset, the model should learn to prioritize the semantic similarity between \textit{A} and \textit{B} over the syntax similarity between \textit{A} and \textit{C}, as shown in part 3.

In particular, we employ Contrastive Learning with the triplet loss function \cite{florian2015triplet}. The input to this function is a triplet of Java classes: \textit{anchor}, \textit{positive}, and \textit{negative}. The \textit{anchor} and \textit{positive} belong to the same microservice while the \textit{negative} class belongs to a different microservice. Using \textit{hard negatives} in triplet CL has been shown to enhance the training \cite{florian2015triplet} where \textit{hard negatives} refer to samples that are similar to the \textit{anchor} but should not be regrouped together. This is why the \textit{negative}, in our case, must belong to the same application. In fact, source code samples from the same application often share structural and syntactical similarities. Thus, using them as \textit{hard negatives} encourages the model to focus on the semantics related to microservices and business logic instead. The triplet loss is defined as follows: 
\begin{equation}\label{eq:tripletloss}
L(a, p, n) = \max(0, \|a - p\|_2 - \|a - n\|_2 + \alpha)
\end{equation}
Where \(a\), \(p\) and \(n\) are the embeddings of the \textit{anchor}, \textit{positive} and \textit{negative} samples. \(\alpha\) is the margin between \textit{positive} and \textit{negative} pairs and \(\|\cdot\|_2\) denotes the Euclidean norm.

For the smaller models (e.g. UnixCoder), we train all of their weights. As for LLMs, we employ the Low-Rank Adaptation (LoRA) method \cite{edward2022lora} to significantly reduce the computational cost and memory requirements while keeping a competitive performance. LoRA has been utilized by recent state-of-the-art models \cite{parishad2024llm2vec,lee2024nvembed}. As for DT LLMs, we adopted the following instruction to enhance model performance \cite{muennighoff2024gritlm,lee2024nvembed,parishad2024llm2vec}:

\begin{center}
\resizebox{\columnwidth}{!}{
\texttt{Given the source code, retrieve the bounded contexts;}}
\end{center}

% \subsection{Creating the Datasets}
\subsection{Creating the Dataset}\label{subsec:dataset}

One of the biggest challenges in microservices decomposition research is the lack of data for supervised learning solutions \cite{oumoussa2024decompsurvey,Abgaz2023decompsurvey}. Monoliths rarely have a single definitive decomposition, making it difficult to rely on existing decompositions for model fine-tuning. However, the open-source ecosystem is abundant with Microservices solutions with various frameworks and diverse designs. We can leverage this fact to construct a dataset for training and evaluating the EMs which is only possible due to LLMs' generalization ability and Contrastive Learning self-supervised methods. The dataset creation process involves three steps:

\subsubsection{Application selection} focuses on choosing microservices repositories, primarily Java applications. Sources include a curated list \cite{rahman2019curatedset}, applications from decomposition research \cite{khaled2022hierdecomp,kalia2021mono2micro,desai2021cogcn,faustino2022stepwise}, and Github API queries. The latter includes only repositories with at least 10 Github stars, that have Java source files and that verify this regex:
\begin{center}
\resizebox{\columnwidth}{!}{\texttt{micro( |-)?services?( |-)(architecture|system|application)}}
\end{center}

\subsubsection{Repository Analysis} involves cloning and analyzing the selected repositories to filter out false positives. We define the sub-directory pattern \textit{"main/java"} as a microservice source root and remove projects with less than 2 microservices. For the remaining projects, we extract their classes and match their source code samples with the corresponding microservice.
\subsubsection{Triplets Sampling} creates the training dataset based on the extracted classes and microservices. Let  \( K \) be the maximum number of samples. We iteratively select random samples to create anchor, positive, and negative triplets. Anchor and positive classes are chosen from the same microservice within a randomly selected repository, while the negative class comes from a different service in the same repository. This process repeats  \( K \) times, after which duplicates are removed.

\subsection{Generating the Decompositions}\label{subsec:inference}
The \textbf{Inference} component in MonoEmbed is responsible for generating decompositions using the embeddings from the \textbf{Analysis} component. By relying on advanced LLMs and our fine-tuning process, the high-level and efficient embeddings will enable better performance while reducing the complexity of the inference. While more advanced approaches such as GNNs can be used for this component, the current implementation of this phase will focus on clustering algorithms. In fact, these algorithms often rely on distances and similarities between the feature vectors of their inputs. Which is why EMs that generate high quality embeddings are well suited with clustering algorithms. In addition, we employ a normalization step, using the z-score standardization method to scale the feature values within the embeddings matrix. This step helps mitigate the impact of variance between feature values.

The output of the \textbf{Inference} component is a vector of integer values which specifies the suggested microservices decomposition. As such, a decomposition is defined as $D = [M_1, M_2, ..., M_K]$. Each microservice in the decomposition is defined as a subset of classes  $M_i = \{c_j | j \in [0, N]\}$ and $|M_i| < N$ where $N$ is the number of classes.

\section{Evaluation}
\subsection{Research Questions}
% In the evaluation phase, we aim to answer the following research questions:
\begin{itemize}
    \item \textbf{RQ1}: How do Pre-Trained models perform in representing classes for the purpose of the decomposition problem?
    \item \textbf{RQ2}:
    \begin{itemize}
        \item \textbf{RQ2-A}: Does our fine-tuning process improve the performance of the Pre-Trained models? \item \textbf{RQ2-B}: How does the size of the training samples impact the performance of the fine-tuning?
    \end{itemize}
    \item \textbf{RQ3}: Which clustering algorithms should be used with our analysis model?
    \item \textbf{RQ4}: How does our approach perform when compared with decomposition benchmarks?
\end{itemize}

% \subsection{Pre-Trained Models}
\subsection{RQ1: The Quality of Pre-Trained Model Embeddings}
\subsubsection{Experimental Setup}
In this phase, we evaluate potential embedding models in the context of the decomposition problem. More specifically, we evaluate the quality of the vectors generated by the models in Table \ref{tab:ptmodels}, where the closer a microservice’s classes’ vectors are, the better the representation is. We include the 4 static analysis representations from section \ref{subsec:mexample} as benchmarks.

\subsubsection{Implementation Details}
We generated class-level embedding vectors for the evaluation applications as follows: \textit{Code2Vec} \cite{alon2018code2vec} and \textit{Code2Seq} \cite{alon2019code2seq} can generate embeddings only at the method-level which is why we define their class-level embeddings as the mean of their corresponding methods \cite{aldebagy2021code2vec}. For the rest, we generate the class embeddings directly. We used the 512-context Java version of \textit{CuBERT}. For \textit{CodeBERT}, \textit{GraphCodeBERT}, and \textit{UnixCoder}, we selected the best-performing input modality based on the mean evaluation scores. We used the last token pooling method for generating sequence embeddings in the case of DT LLMs. \textit{GritLM}, \textit{NVEmbed}, and \textit{LLM2Vec} embeddings were generated based on the authors' shared code and models. For \textit{LLM2Vec}, we used the supervised learning version \cite{parishad2024llm2vecsupervisedmodelcard} of MNTP-LLama3 \cite{aimeta2024llama3modelcard} fine-tune. For the CEM providers, we report results from the best-performing model.

\subsubsection{Evaluation Dataset}\label{subsubsec:rq1dataset}
We select a dataset of 6 microservices and modular applications with varying scales as shown in Table \ref{tab:microapps}. The applications Spring Petclinic Microservices \cite{microapps2024petclinic}, Acme Air \cite{monoapps2024acmeair}, Eventuate Kanban Board \cite{microapps2024kanban} and Event Sourcing \cite{microapps2024eventsourcing} were selected due to their usage in decomposition approaches \cite{kalia2021mono2micro,mathai2022chgnn,khaled2022msextractor,khaled2022hydecomp}. We include the microservices \cite{microapps2024socialeditionmicroservices}  and modular monolith \cite{microapps2024socialeditionmodular} versions of the monolithic application Social Edition which has been used in the decomposition context \cite{faustino2022stepwise}.

% to bridge the gap between microservices and monolith inputs.

\begin{table}[ht]
\caption{Microservices and Modular Monolith application data.}
\label{tab:microapps}
\resizebox{\columnwidth}{!}{%
\begin{tabular}{@{}lllll@{}}
\toprule
Application                      & Short Name & \# of classes & \# of methods & \# of microservices \\ \midrule
PetClinic Microservices    & PetClinic  & 36            & 52           & 7                   \\
Acme Air                 & AcmeAir  & 86           & 758          & 7                  \\
Eventuate Kanban Board                 & ESKanban  & 112           & 659          & 19                  \\
Microservices Event Sourcing & CESource  & 119           & 648          & 12                  \\
Social Edition Modular       & SEMod  & 376           & 3175          & 9                   \\
Social Edition Microservices   & SEMicro  & 509           & 4224         & 9                   \\ \bottomrule
\end{tabular}
}
\end{table}

\subsubsection{Evaluation Metrics}\label{subsubsec:rq1metrics}
For any application, the closer the embeddings of classes that should belong to the same microservices are and the farther away those of different microservices are from each other the better the quality of the embedding model. We present in Algorithm \ref{alg:embscore} the pseudo-code for measuring a score that can quantify this quality. We calculate the Embedding Quality Score across all evaluation repositories. For each application, the algorithm extracts classes and their associated microservices, generates standardized embeddings of their source code using the given model, and gets all unique permutations of these embeddings. Each permutation corresponds to a label where 1 means classes are in the same microservice and 0 if not. It then computes cosine similarities between the pairs, normalizes them, and calculates a balanced binary cross-entropy loss score comparing the similarities to the actual microservice associations. We chose the balanced loss function instead of the standard log loss function since the number of labels 0 significantly outnumbers label 1. The algorithm returns both the mean score across all repositories and a list of individual repository scores which we will use for evaluating the model. The lower the score is, the closer the embeddings to the actual microservices' structure and the better their quality is.

\begin{algorithm}[htbp]
\caption{Embedding Quality Score}\label{alg:embscore}
\SetKwInOut{Input}{Input}\SetKwInOut{Output}{Output}
\Input{model, repositories}
\Output{meanScore, allScores}
\BlankLine
\For{repo in repositories}{
    code, services = extractClasses(repo)\\
    embeddings = standarize(model.encode(code))\\
    couples, labels = toPairwise(embeddings, services)\\
    similarities = minMax(cosine(couples))\\
    allScores.add(balancedLogLoss(similarities, labels))\\
}
\Return mean(allScores), allScores
\end{algorithm}

\subsubsection{Results}
The Table \ref{tab:rq1_results} showcases the evaluation results for the Pre-Trained models we considered on the evaluation applications and their mean score. The models are sorted by their mean score from lowest (best) to highest (worst). We highlight the 3 best models in each application and the mean column with bold, double underline and simple underline. Overall, the mean score results in Table \ref{tab:rq1_results} demonstrate a similar pattern to the model rankings in MTEB\cref{footnote:mteb}. The first 8 ranks in our benchmark were achieved by LLMs and CEMs showcasing their adaptability. In fact, \textit{VoyageAI}, had the best score while the other 2 CEMs reached the $6^{th}$ and $7^{th}$ ranks. Nonetheless, all LMEs managed to perform well in comparison where \textit{NVEmbed} achieved the second lowest mean score and \textit{GritLM} and \textit{LLM2Vec} had better or similar scores to the other CEMs. However, most DT LLMs underachieved in comparison and were surpassed by much smaller ETs and benchmarks, with the exception of \textit{SFR Mistral} and \textit{E5 Mistral} which landed the $3^{rd}$ and $4^{th}$ ranks. Unlike the rest of the DTs, these models were fine-tuned for representation learning which explains their performance. \textit{DeepSeek Coder} and \textit{CodeLLama}, code specialized DT LLMs, scored worse than most models but better than \textit{Llama-3}, a generalist LLM, indicating that training objectives and model architecture outweigh domain specialization. However, most ETs performed worse than LLMs and even benchmarks like \textit{ST-Interactions}. The unified models \textit{UnixCoder} and \textit{CodeT5+} are an exception as they managed to land the $9^{th}$ and $10^{th}$ place, thus competing with the CEMs and LMEs. The final ranks were occupied by the traditional Embedding Models \textit{Code2Vec} and \textit{Code2Seq} ($20^{th}$ and $21^{st}$) and \textit{ST-Calls}. The method-level analysis of these models is a likely reason for their performance since they can struggle with annotation-based applications such as \textit{PetClinic} where most classes are data models. While \textit{ST-Interactions} achieved the $11^{th}$, beating most models we considered and even had the best score for PetClinic, \textit{ST-Calls} had significantly higher scores on all applications. We believe that this difference is due to the sparse nature of the dependency matrix in \textit{ST-Calls} unlike \textit{ST-Interactions}. However, we would like to note that applying static analysis on Microservices applications does not reflect their performance on Monoliths due to the lack of inter-microservices interactions. The lack of misleading dependencies can lead to better scores (\textit{ST-Interactions}) while the sparsity can lead to worse scores (\textit{ST-Calls}).

\begin{tcolorbox}[colback=gray!10!white, colframe=gray!90!black]
\textbf{Summary:} We show that LLM based encoders, like VoyageAI, NVEmbed and SFR Mistral, generate better representation vectors in the decomposition context than the methods employed in related works (e.g. static analysis, Code2Vec, CodeBERT). 
\end{tcolorbox}

\begin{table}[h]
    \centering
    \caption{Embedding model scores on the evaluation applications.}
    \label{tab:rq1_results}
    \resizebox{\linewidth}{!}{%

\begin{tabular}{ll|lllllll}
\toprule
{} &            Model &                Mean &           PetClinic &             AcmeAir &            ESKanban &            CESource &             SEMicro &               SEMod \\
\midrule
1  &         VoyageAI &     \textbf{0.6387} &              0.6900 &     \textbf{0.5944} &     \uuline{0.5752} &              0.6678 &     \textbf{0.6684} &     \textbf{0.6364} \\
2  &          NVEmbed &     \uuline{0.6405} &              0.6833 &     \uuline{0.6150} &     \textbf{0.5710} &  \underline{0.6617} &  \underline{0.6712} &     \uuline{0.6409} \\
3  &      SFR Mistral &  \underline{0.6409} &     \uuline{0.6510} &              0.6360 &  \underline{0.5787} &     \textbf{0.6588} &              0.6765 &              0.6445 \\
4  &       E5 Mistral &              0.6417 &  \underline{0.6521} &              0.6360 &              0.5814 &     \uuline{0.6596} &              0.6764 &              0.6449 \\
5  &           GritLM &              0.6465 &              0.6670 &              0.6270 &              0.5970 &              0.6660 &              0.6737 &              0.6482 \\
6  &           OpenAI &              0.6482 &              0.6897 &  \underline{0.6215} &              0.5850 &              0.6798 &     \uuline{0.6702} &  \underline{0.6429} \\
7  &           Cohere &              0.6506 &              0.6666 &              0.6285 &              0.5920 &              0.6779 &              0.6934 &              0.6450 \\
8  &          LLM2Vec &              0.6507 &              0.6717 &              0.6433 &              0.5995 &              0.6651 &              0.6738 &              0.6510 \\
9  &        UnixCoder &              0.6560 &              0.6694 &              0.6661 &              0.5955 &              0.6739 &              0.6796 &              0.6514 \\
10 &          CodeT5+ &              0.6562 &              0.6848 &              0.6523 &              0.5989 &              0.6790 &              0.6716 &              0.6508 \\
11 &  ST-Interactions &              0.6651 &     \textbf{0.6430} &              0.6687 &              0.6455 &              0.6688 &              0.6843 &              0.6801 \\
12 &    GraphCodeBERT &              0.6728 &              0.6978 &              0.6514 &              0.6278 &              0.7041 &              0.6884 &              0.6669 \\
13 &           SM-BoW &              0.7070 &              0.7316 &              0.7412 &              0.6295 &              0.7227 &              0.7405 &              0.6765 \\
14 &      DeepSeek Coder   &              0.7104 &              0.7225 &              0.6975 &              0.6785 &              0.7239 &              0.7253 &              0.7143 \\
15 &         SM-TFIDF &              0.7122 &              0.7444 &              0.7156 &              0.6608 &              0.7200 &              0.7393 &              0.6933 \\
16 &   CodeLlama &              0.7138 &              0.7171 &              0.6970 &              0.6801 &              0.7403 &              0.7288 &              0.7192 \\
17 &          Llama-3 &              0.7240 &              0.7076 &              0.7108 &              0.7110 &              0.7408 &              0.7412 &              0.7329 \\
18 &         CodeBERT &              0.7365 &              0.7984 &              0.6790 &              0.7108 &              0.7662 &              0.7425 &              0.7221 \\
19 &           CuBERT &              0.8121 &              0.8243 &              0.7571 &              0.8127 &              0.8472 &              0.8398 &              0.7913 \\
20 &         Code2Vec &              1.3176 &              3.4025 &              0.9142 &              0.8453 &              1.1773 &              0.8365 &              0.7299 \\
21 &         Code2Seq &              1.4238 &              3.4787 &              1.0459 &              0.9711 &              1.3216 &              0.8796 &              0.8460 \\
22 &         ST-Calls &              7.6066 &             12.4624 &              9.7032 &              8.6021 &              8.3805 &              3.8777 &              2.6134 \\
\bottomrule
\end{tabular}

    }

\end{table}

% \subsection{Fine-Tuned Models}\label{subsec:ftmodels}
\subsection{RQ2: The Quality of Fine-Tuned Model Embeddings}\label{subsec:ftmodels}
\subsubsection{Experimental Setup}

For this research question, we would like to, first, evaluate how much are we able to improve the Pre-Trained models through fine-tuning and, second, to identify the impact of the scale of the training set data samples on the quality of the embeddings. For fine-tuning the models, we generate a dataset as described in section \ref{subsec:finetune}. We exclude all of the evaluation applications described in Table \ref{tab:microapps} from the training dataset. The generated dataset has approximately 340K unique triplet samples. For the evaluation, we used the same score and applications described in section \ref{subsubsec:rq1metrics} and section \ref{subsubsec:rq1dataset}. We present the score for each application and their mean where the 3 best models are highlighted with bold text, double underline and simple underline.

\subsubsection{Implementation Details}

In addition to the process described in section \ref{subsec:finetune}, we provide the implementation details for the chosen models: For \textit{UnixCoder} and \textit{CodeT5+}, we used the recommended learning rate value of 2e-5 \cite{chi2019finetunebert} for fine-tuning the models.  For the rest of the models, we train them using the LoRA method due to their sizes. In the case of \textit{LLM2Vec}, we fine-tune the MNTP Llama 3 version of the proposed model \cite{parishad2024llm2vecmntpmodelcard} instead of the supervised version \cite{parishad2024llm2vecsupervisedmodelcard} based on the authors' training observations \cite{parishad2024llm2vec}. We used the same setup and hyper-parameters they suggest\footnote{\url{https://github.com/McGill-NLP/llm2vec}}. For \textit{GritLM}, we used the author's configuration and hyper-parameters\footnote{\url{https://github.com/ContextualAI/gritlm}}. As for \textit{SFR-Mistral}, we add a "lasttoken" pooling layer to the model and used the same setup as \textit{LLM2Vec} for the training.

\subsubsection{Base Models Comparison (RQ2-A)}
% for new line
\phantom{1}

\textbf{Motivation}: For the first objective in this research question, we fine-tune 5 models among the 10 models that achieved the best mean score in Table \ref{tab:appscores}: \textit{SFR-Mistral}, \textit{GritLM}, \textit{UnixCoder}, \textit{CodeT5+} and \textit{LLM2Vec}. The selection of these models relied on their availability, their category and our available resources. All of the models were trained on 10\% of the dataset (34K samples). \textit{SFR-Mistral}, \textit{UnixCoder} and \textit{CodeT5+} were chosen because they achieved the best scores among their respective groups. We select \textit{GritLM} to represent the group of LMEs since we were not able to fine-tune \textit{NVEmbed} on our resources with the same training setup as the rest of the models. \textit{GritLM} is a unified model (i.e. it has been trained for both embedding and generative tasks) while \textit{LLM2Vec}, through its modular 3-step tuning process, is specialized for embeddings. For this reason, we decided to consider \textit{LLM2Vec} as well.

% For this reason, \textit{LLM2Vec} and due to its modular 3-step tuning process, we decided to consider \textit{LLM2Vec} as well.

\textbf{Results}: Table \ref{tab:ftscores_models} shows the evaluation results of the Fine-Tuned (FT) Models. We include the results of the Pre-Trained model \textit{VoyageAI} as a reference. The models are sorted based on their average performance. We can observe in the table that all of the models managed to outperform the benchmark \textit{VoyageAI} on all of the evaluation applications with the exception of the application \textit{AcmeAir} and \textit{ME-gritlm} and \textit{ME-codet5+} in \textit{ESKanban}. In fact, there is a 0.075 reduction in score between \textit{ME-llm2vec} and the benchmark. Both \textit{ME-llm2vec} and \textit{ME-SFR-Mistral} had the best mean scores and outperformed the rest on most applications with \textit{ME-llm2vec} having a slight edge. \textit{ME-gritlm}, unlike the other LLMs, was surpassed by \textit{ME-unixcoder} showcasing the advantage of using a purely representation learning focused LLM like \textit{LLM2Vec}.

\begin{table}[]

    \centering
    \caption{Fine-tuned models' scores on the evaluation applications. }
    \label{tab:ftscores_models}
    \resizebox{\linewidth}{!}{%

   \begin{tabular}{l|lllllll}
\toprule
         Model &               Mean &          PetClinic &            AcmeAir &           ESKanban &           CESource &            SEMicro &              SEMod \\
\midrule
    ME-llm2vec &    \textbf{0.5626} &    \textbf{0.5499} &    \uuline{0.6008} &    \textbf{0.4870} &    \textbf{0.6159} & \underline{0.5819} & \underline{0.5401} \\
ME-SFR-Mistral &    \uuline{0.5666} &    \uuline{0.5579} & \underline{0.6215} &    \uuline{0.4943} &    \uuline{0.6254} &    \uuline{0.5757} &    \textbf{0.5247} \\
  ME-unixcoder & \underline{0.5968} & \underline{0.5895} &             0.6386 & \underline{0.5646} &             0.6315 &             0.5978 &             0.5586 \\
     ME-gritlm &             0.5999 &             0.5970 &             0.6612 &             0.6084 &             0.6506 &    \textbf{0.5486} &    \uuline{0.5337} \\
    ME-codet5+ &             0.6071 &             0.6009 &             0.6502 &             0.6020 & \underline{0.6256} &             0.5955 &             0.5687 \\
      VoyageAI &             0.6387 &             0.6900 &    \textbf{0.5944} &             0.5752 &             0.6678 &             0.6684 &             0.6364 \\
\bottomrule
\end{tabular}

}
\end{table}

\subsubsection{Sample Size Comparison (RQ2-B)}
% for new line
\phantom{1}

\textbf{Motivation}: For the second objective, we train each of the models \textit{UnixCoder}, \textit{CodeT5+} and \textit{LLM2Vec} on different numbers of samples: 3K, 34K and 340K samples ( 1\%, 10\% and 100\% of the generated dataset). We evaluate the impact of augmenting the dataset from the perspective of the number of samples without modifying the number of repositories.

\textbf{Results}: Table \ref{tab:ftscores_samples} shows the quality scores of the fine-tuned models on different scales of training samples. The results show that all of the models were able to surpass \textit{VoyageAI}'s score in \ref{tab:ftscores_models} even at only 3K training samples. In fact, the model \textit{ME-llm2vec} with 3K samples achieved better scores than both \textit{ME-unixcoder} and \textit{ME-codet5+} when trained on 340K samples. By increasing the scale of the dataset, we can observe that \textit{ME-llm2vec} continues to improve. However, transitioning from 34K to 340K samples had a less significant effect than the transition from 3K to 34K. This result showcases the diminishing returns of scaling up the number of samples. Improving the model's performance even further will likely require increasing the number of repositories in the dataset or incorporating a more complex sample selection process. Notably, the \textit{ME-llm2vec} that was trained with 340K samples surpassed all of the models in Table \ref{tab:ftscores_models}. While both \textit{ME-unixcoder} and \textit{ME-codet5+} did benefit from more samples, \textit{ME-unixcoder} improvement rate was higher than \textit{ME-codet5+}'s, and thus switching their ranking after 34K samples.

% \begin{tcolorbox}[colback=gray!10!white, colframe=gray!90!black]
% \textbf{Summary:} Even with a small sample size, we can adapt the models to the decomposition problem and surpass Pre-Trained models. Increasing the training data scale will continue to improve these models further but at an exponentially decreasing rate. 
% \end{tcolorbox} 
% all of our fine-tuned models managed to surpass the Pre-Trained models, with \textit{ME-llm2vec} and \textit{ME-SFR-Mistral} achieving significantly better score.
\begin{tcolorbox}[colback=gray!10!white, colframe=gray!90!black]
\textbf{Summary:} RQ2-A shows that all of our models managed to surpass the Pre-Trained models, with \textit{ME-llm2vec} achieving a significantly better score while RQ2-B shows that it is achievable even with a small sample size. Increasing the scale further will improve these models but at an exponentially decreasing rate. 
\end{tcolorbox} 

\begin{table}[]

    \centering
    \caption{Fine-tuned models' scores per number of training samples.}
    \label{tab:ftscores_samples}
    \resizebox{\linewidth}{!}{%

    \begin{tabular}{l|l|lllllll}
\toprule
\makecell{Samples} &  \makecell{Model} & \makecell{Mean} & \makecell{PetClinic} & \makecell{AcmeAir} & \makecell{ESKanban} & \makecell{CESource} & \makecell{SEMicro} & \makecell{SEMod} \\
\midrule
   &   ME-llm2vec &    \textbf{0.5483} &    \textbf{0.5386} &    \textbf{0.5932} &    \uuline{0.5009} &    \uuline{0.6211} &    \textbf{0.5440} &    \textbf{0.4922} \\
   \makecell{340K} & ME-unixcoder &             0.5881 &             0.5973 &             0.6633 &             0.5393 &             0.6293 & \underline{0.5792} & \underline{0.5199} \\
   &   ME-codet5+ &             0.5956 &             0.5648 &             0.6827 &             0.5583 &             0.6678 &             0.5931 &    \uuline{0.5067} \\
\midrule
    &   ME-llm2vec &    \uuline{0.5626} &    \uuline{0.5499} &    \uuline{0.6008} &    \textbf{0.4870} &    \textbf{0.6159} &             0.5819 &             0.5401 \\
    \makecell{34K} & ME-unixcoder &             0.5968 &             0.5895 &             0.6386 &             0.5646 &             0.6315 &             0.5978 &             0.5586 \\
    &   ME-codet5+ &             0.6071 &             0.6009 &             0.6502 &             0.6020 & \underline{0.6256} &             0.5955 &             0.5687 \\
\midrule
     &   ME-llm2vec & \underline{0.5851} & \underline{0.5639} & \underline{0.6054} & \underline{0.5259} &             0.6275 &             0.6116 &             0.5764 \\
     \makecell{3K} &   ME-codet5+ &             0.6079 &             0.6141 &             0.6746 &             0.5904 &             0.6455 &    \uuline{0.5751} &             0.5475 \\
      & ME-unixcoder &             0.6195 &             0.6184 &             0.6596 &             0.5364 &             0.6514 &             0.6462 &             0.6051 \\
\bottomrule
\end{tabular}

}
\end{table}

% \subsection{Clustering Algorithms}
\subsection{RQ3: The Clustering Algorithms}\label{subsec:rq3clustering}

\begin{figure}[htbp]
\centering
\includegraphics[width=\linewidth]{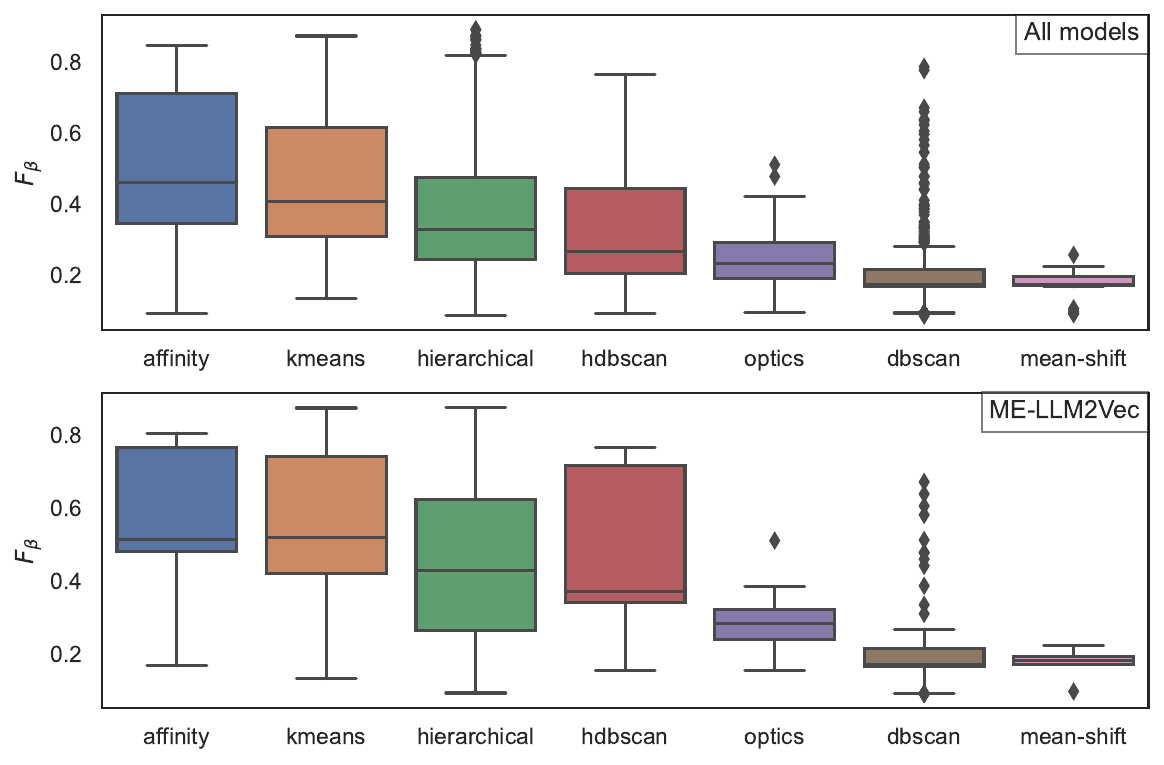}
\caption{Evaluation $F_\beta$ scores for each clustering algorithm.} \label{fig:rq3boxplots}
\end{figure}

\subsubsection{Experimental Setup}
We experiment with different clustering algorithms and evaluate their general performance on the decomposition problem when used in conjunction with different LMs and across different input configurations. We select 7 popular clustering algorithms and we chose multiple values for their most impactful hyper-parameters. Using this setup, we generate multiple decompositions with the FT-Models: \textit{ME-llm2vec-340K}, \textit{ME-SFR-Mistral}, \textit{ME-gritlm}, \textit{ME-codet5-340K} and \textit{ME-unixcoder-340K} on the applications from Table \ref{tab:microapps}. We include, as well, the benchmarks VoyageAI and NVEmbed.

\subsubsection{Metrics}
To evaluate the clustering algorithms in the context of our problem, we need to use metrics that can quantify how close is the generated decomposition from the actual microservices. For this reason, we rely on the relationship between the classes and the microservices they belong to. For an application, for which we know the true microservices, we can evaluate a decomposition by transforming it into a binary vector of unique class permutations. Each value represents whether the corresponding class couple is in the same microservice or not. This vector represents the "predictions" while the "labels" can be extracted using the same approach on the true partitioning. The evaluation score, afterwards, is measured by applying the $F_\beta$ formula on the predictions and labels vectors in order to account for both precision and recall. $F_\beta$, with a $\beta$ value equal to 0.25, was chosen over the more commonly used $F_1$ score because of the unbalanced nature of the relationship vectors \cite{khaled2024rldec}. Higher values of $F_\beta$ are better.

\subsubsection{Results}

Fig. \ref{fig:rq3boxplots} presents $F_\beta$ scores for various clustering algorithms with all models and with \textit{ME-llm2vec-340K} results. For All models, \textit{Affinity Propagation} \cite{brenden2007affinity} achieved the highest median $F_\beta$, followed closely by \textit{K-Means}, which showed higher maximum and minimum values. \textit{Hierarchical} clustering \cite{ward1963hierarchical}, despite having a lower median score, achieved the highest $F_\beta$. Both \textit{HDSCAN} \cite{ricardo2013hdbscan} and \textit{OPTICS} \cite{ankerst1999optics} had better scores than the algorithm they extend, \textit{DBSCAN} \cite{ester1996dbscan}, with \textit{HDSCAN} reaching a notably higher maximum than \textit{OPTICS}. Mean-Shift \cite{comaniciu2002meanshift}, however, had the lowest scores. In Fig. \ref{fig:rq3boxplots} for \textit{ME-llm2vec}, most algorithms showed higher median scores while their order is intact. Notably, \textit{K-Means}' median score increased significantly, approaching that of \textit{Affinity Propagation}.

\textit{K-Means} and \textit{Hierarchical} have achieved high maximum scores making them equally valid choices if the target number of microservices is known. However, \textit{Affinity Propagation} scored higher on average making it a better default option. Although \textit{HDSCAN} is less dependent on hyper-parameters, its low median score shows that it is not as consistent. Inspecting the hyper-parameter details further, we observed that damping values in the range (0.5, 0.8) for \textit{Affinity Propagation} provide a consistent score which can drop when above this threshold. 

% As for similarity-based algorithms, like \textit{HDSCAN}, we observed that the cosine similarity provides better results on average.

\begin{tcolorbox}[colback=gray!10!white, colframe=gray!90!black]
\textbf{Summary:} Affinity Propagation provides on average better results while K-Means and Hierarchical have higher scores when given the number of microservices.
\end{tcolorbox}

% \subsection{Comparison with the Benchmarks}
\subsection{RQ4: Comparison with the Decomposition Benchmarks}\label{subsec:benchmarks}
\subsubsection{Experimental Setup}
In this RQ, we evaluate the performance of our approach on monolithic applications and compare it with benchmark decomposition methods. We select 8 monolithic applications with varying numbers of classes and that have been used in decomposition problems \cite{khaled2022msextractor,kalia2021mono2micro,jin2021fosci}. We apply our approach and 6 additional benchmarks on these applications and measure 7 evaluation metrics that represent different qualities in the generated microservices. The Table \ref{tab:monoapps} shows the details of the monolithic applications

The benchmark decomposition approaches that we compare with are: \textit{Code2VecDec} \cite{aldebagy2021code2vec}, \textit{CHGNN} \cite{mathai2022chgnn}, \textit{Deeply} \cite{yedida2023deeply}, \textit{Mono2Micro} \cite{kalia2021mono2micro}, \textit{MSExtractor} \cite{khaled2022msextractor} and \textit{TopicDec} \cite{brito2021topicmodeling}. Since some of the approaches, such as \textit{TopicDec} and \textit{Mono2Micro}, generate multiple decompositions or do not have predefined hyper-parameters, We generate multiple results for the rest of the approaches by varying their potential hyper-parameters. \textit{Deeply} and \textit{Code2VecDec} are the exceptions since the former always generates a single decomposition while the latter’s authors specified the recommended hyper-parameters.  We apply these approaches on all of the applications except for \textit{CHGNN} due to its requirement of manually defined input seeds. As such, we evaluate \textit{CHGNN} only on Plants \cite{monoapps2024plants}, AcmeAir \cite{monoapps2024acmeair} and DayTrader \cite{monoapps2024daytrader} for which the authors provide input data. As for \textit{MonoEmbed}, we used \textit{ME-llm2vec-340K} due to its performance in Table \ref{tab:ftscores_samples} and Affinity Propagation based on our observation in section \ref{subsec:rq3clustering}.

\begin{table}[h]
\caption{Monolithic application data.}\label{tab:monoapps}
\resizebox{\columnwidth}{!}{%
\begin{tabular}{@{}lllll@{}}
\toprule
Application   & \# of classes & \# of methods & \# of semantic terms & \# of unique calls \\ \midrule
Plants \cite{monoapps2024plants}    & 40            & 1109          & 607                  & 123                \\
 JPetStore \cite{monoapps2024jpetstore} & 43& 876& 231&96\\
PetClinic \cite{monoapps2024petcliniclegacy} & 60            & 987           & 475                  & 106                \\
ACMEair \cite{monoapps2024acmeair}   & 86            & 1629          & 553                  & 123                \\
 PartsMRP \cite{monoapps2024partsunlimited} & 100& 1951& 423&234\\
DayTrader \cite{monoapps2024daytrader} & 118           & 2577          & 607                  & 282                \\
 7ep-demo \cite{monoapps20247epdemo} & 119& 2702& 1048&204\\
 Roller \cite{monoapps2024roller}   & 531           & 15426         & 2632                 & 2195               \\  \bottomrule
\end{tabular}
}
\end{table}

\subsubsection{Evaluation Metrics}
The evaluation in this RQ is based on monolithic applications for which a true and correct decomposition rarely exists. As such, we utilized 6 metrics from related works and an aggregate score that measure certain qualities in the decomposition. CoHesion at the Message level (CHM)\cite{jin2021fosci,athanasopoulos2015cohesion} is commonly used to measure the structural cohesiveness of the components in the microservices. The assumption in this case is that the better the structural similarity is the better the decomposition is. CoHesion at the Domain level (CHD)\cite{jin2021fosci,athanasopoulos2015cohesion} is similar to CHM but focuses on the domain instead. Business Case Purity (BCP)\cite{kalia2021mono2micro} quantifies the cohesion from a business usecase standpoint. Inter-Call Percentage (ICP)\cite{kalia2021mono2micro} is a measurement of the coupling between the microservices. Non-Extreme Distribution (NED)\cite{desai2021cogcn} reflects the size of the microservices penalizing extremely small or large microservices. Coverage (COV) is the percentage of classes that were included in the decomposition.

We define an aggregate score (SCORE) combining all metrics, inspired by \textit{CHGNN}'s evaluation \cite{mathai2022chgnn}. This approach addresses two challenges: Metric scales vary by application (e.g., median BCP for JPetStore is 2.6, while for PetClinic it's 0.8) and and vary between each other. For a fair comparison:
\begin{enumerate}

\item We standardize evaluation metrics for each application.
\item We combine scaled metrics using a weighted sum:
\begin{itemize}

\item CHM and CHD (to be maximized): weight = 2
\item BCP and ICP (to be minimized): weight = -2
\item NED and COV (secondary qualities): weights = -1 and 1 respectively
\end{itemize}
\end{enumerate}

A higher SCORE indicates a more balanced decomposition. In addition, this scaling allows aggregation across applications.

\begin{table}[h]

    \centering
    \caption{Scaled evaluation metrics for each decomposition approach. }
    \label{tab:monoscores}
    \resizebox{\linewidth}{!}{%

\begin{tabular}{l|lllllll}
\toprule
   Approach &     CHM$\nearrow$ &     CHD$\nearrow$ &      BCP$\searrow$ &      ICP$\searrow$ &      NED$\searrow$ &     COV$\nearrow$ &   SCORE$\nearrow$ \\
\midrule
  MonoEmbed &    \textbf{1.168} &    \textbf{1.375} &             -0.527 &              0.323 &              0.252 &    \textbf{0.818} &    \textbf{6.061} \\
 Mono2Micro & \underline{0.174} &    \uuline{0.317} &    \textbf{-1.210} & \underline{-0.719} &              0.231 &            -1.124 &    \uuline{3.485} \\
Code2VecDec &    \uuline{0.461} & \underline{0.187} &    \uuline{-0.665} &              1.144 & \underline{-0.091} & \underline{0.678} & \underline{1.107} \\
MSExtractor &             0.068 &             0.022 &              0.227 &              0.543 &              0.215 &    \textbf{0.818} &            -0.757 \\
     Deeply &            -0.782 &            -0.852 & \underline{-0.643} &    \textbf{-1.500} &              0.599 &            -1.219 &            -0.801 \\
   TopicDec &            -0.538 &            -0.368 &             -0.061 &    \uuline{-0.776} &    \uuline{-0.339} &            -1.170 &            -0.969 \\
      CHGNN &             0.014 &            -0.351 &              0.495 &             -0.375 &    \textbf{-0.613} &            -0.739 &            -1.040 \\
\bottomrule
\end{tabular}

}
\end{table}

\subsubsection{Results}

Table \ref{tab:monoscores} shows the mean scaled metric values across evaluation applications for each decomposition approach while Table \ref{tab:appscores} presents the SCORE values for each benchmark and  application. The best 3 values in each metric are in bold, double underline and simple underline. 

In Table \ref{tab:monoscores}, our approach, \textit{MonoEmbed}, achieved the best SCORE, followed by \textit{Mono2Micro} and \textit{Code2VecDec}. In fact, \textit{MonoEmbed} scored highest in CHM, CHD, and COV, with a BCP value close to the second and third models, demonstrating its ability to group cohesive classes across multiple aspects. It even surpassed \textit{CHGNN}, a dynamic analysis approach, and nearly matched \textit{Deeply}'s BCP score, a metric it optimizes. However, \textit{MonoEmbed}'s higher-than-average ICP score suggests increased inter-microservice coupling.

As shown in Table \ref{tab:appscores}, \textit{MonoEmbed} achieved the highest scores for Plants, AcmeAir, PartUnlimited, and Roller, and second-highest for DayTrader and 7EP-Demo but had worse scores on smaller applications. While \textit{MSExtractor} and \textit{Code2VecDec} excelled with smaller applications, they struggled with the rest, suggesting that simpler methods or method-level granularities may suit smaller applications, which have more focused relationships and bounded contexts. \textit{MonoEmbed} on the other hand demonstrated more consistent performance, especially for larger applications. 

Both \textit{Code2VecDec} and \textit{MonoEmbed} use embedding models and clustering. While \textit{Code2VecDec} performed well on PetClinic, its results were inconsistent and generally lower than \textit{MonoEmbed}'s, highlighting the advantage of fine-tuned contextual EMs. However, \textit{Code2VecDec}'s performance suggests some potential in method-level embeddings. Unlike other benchmarks, \textit{MonoEmbed} does not use predefined representations. \textit{CHGNN}, \textit{Deeply}, and \textit{Mono2Micro} rely on runtime interactions, while \textit{MSExtractor} and \textit{TopicDec} use structural and semantic similarities. These approaches may excel in metrics reflecting similar qualities to their inputs. For instance, \textit{Mono2Micro} achieved low BCP and ICP scores but sacrificed coverage and microservice balance. On the other hand, \textit{MonoEmbed}'s model was trained on numerous microservices applications and learns class groupings based on real-world developer practices. This approach likely contributes to its consistent performance across metrics and applications. While all approaches use source code at some level, \textit{MonoEmbed} does not require in-depth analysis or additional input, potentially explaining further its consistency across applications.

\begin{figure*}[]
\centering
\includegraphics[width=0.9\linewidth]{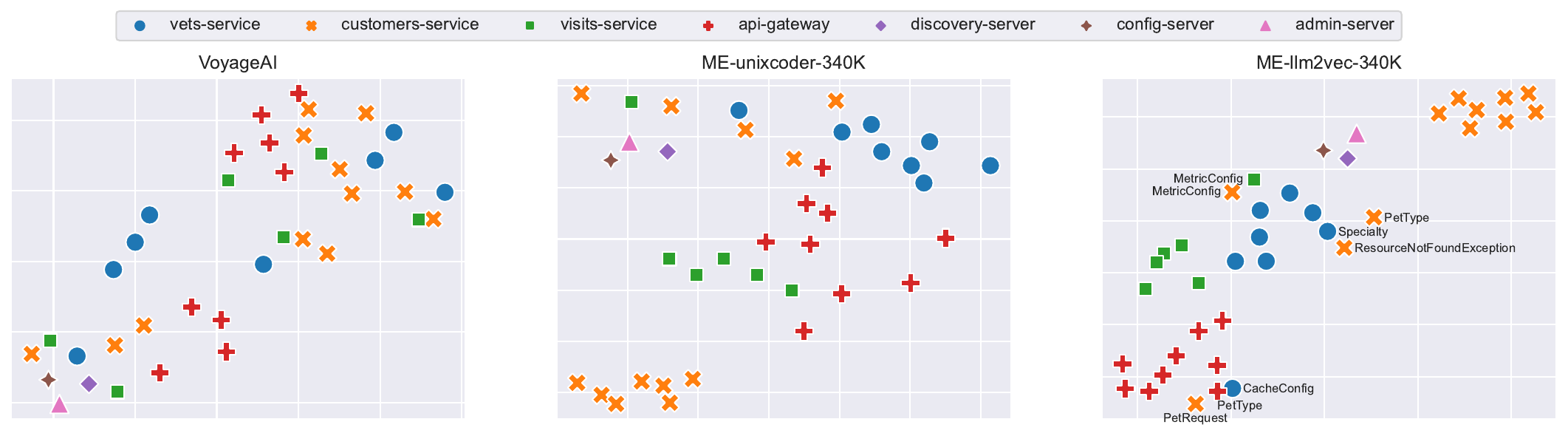}
\caption{A UMAP 2-dimensional projection of VoyageAI, MonoEmbed-unixcoder and MonoEmbed-llm2vec embeddings.} \label{fig:casestudy}
\end{figure*}

\begin{tcolorbox}[colback=gray!10!white, colframe=gray!90!black]
\textbf{Summary:} \textit{MonoEmbed} is able to consistently generate highly cohesive decompositions on different applications and scales while maintaining full coverage, albeit with higher microservice coupling.
\end{tcolorbox}

\begin{table}[]

    \centering
    \caption{Aggregate scores on each monolithic application.}
    \label{tab:appscores}
    \resizebox{\linewidth}{!}{%

    \begin{tabular}{l|lllllll}
\toprule
   Application &       MonoEmbed &         Mono2Micro &        Code2VecDec &        MSExtractor &          Deeply &       TopicDec &  CHGNN \\
\midrule
        Plants &  \textbf{5.986} &             -0.998 &  \underline{0.740} &     \uuline{2.238} &         -10.799 &         -2.791 & -2.359 \\
     JPetStore &          -5.177 &     \uuline{1.668} & \underline{-2.369} &     \textbf{2.339} &          -5.069 &         -3.609 &      - \\
     PetClinic &          -2.069 &             -3.099 &     \textbf{1.209} &  \underline{0.534} &         -10.010 & \uuline{0.892} &      - \\
       AcmeAir &  \textbf{8.521} &     \uuline{4.187} &  \underline{1.999} &             -2.192 &          -1.113 &          0.277 &  0.576 \\
PartsUnlimited & \textbf{14.530} & \underline{-0.661} &             -2.443 &             -1.901 &          -2.789 & \uuline{0.160} &      - \\
     DayTrader & \uuline{11.170} &  \underline{8.606} &              4.382 &             -1.582 & \textbf{20.272} &         -5.415 & -1.336 \\
      7EP-Demo &  \uuline{1.395} &     \textbf{6.038} &             -7.743 & \underline{-0.447} &          -0.659 &         -1.138 &      - \\
        Roller & \textbf{14.129} &  \underline{5.433} &    \uuline{13.082} &             -5.042 &           3.759 &          3.870 &      - \\
\midrule
          Mean &  \textbf{6.061} &     \uuline{3.485} &  \underline{1.107} &             -0.757 &          -0.801 &         -0.969 & -1.040 \\
\bottomrule
\end{tabular}

}
\end{table}

% \section{Discussion and Threats to Validity}
\section{Discussion}
\subsection{Case Study}

Fig. \ref{fig:casestudy} shows 2-D UMAP \cite{mcinnes2020umapuniformmanifoldapproximation} projections of embedding vectors for PetClinic Microservices \cite{microapps2024petclinic} using VoyageAI, ME-unixcoder-340K, and ME-llm2vec-340K. ME-unixcoder and ME-llm2vec show clearer microservice clusters compared to VoyageAI and Fig. \ref{fig:mexample}. For instance, \textit{API-Gateway} service classes are split in VoyageAI's representation but closer in the other models. ME-llm2vec demonstrates the closest grouping within microservices and clearest separation between them. In fact, ME-llm2vec's figure shows only 6 clear outliers which we review in detail and display their names and neighbors.

The two \textit{MetricConfig} classes, \textit{ResourceNotFoundException} and \textit{CacheConfig} lack domain-specific terms since they are utility classes, which highlights the importance of separating them from domain-related ones during the decomposition. However, ME-llm2vec was able correctly represent classes with even slight domain hints. For instance, most models struggle to differentiate between the nearly identical entry-point classes (e.g. \textit{ConfigServerApplication}), as seen in Fig. \ref{fig:mexample} and \ref{fig:casestudy} while ME-llm2vec managed to correctly place them within their services. On the other hand, the class \textit{PetRequest}, which was closer to \textit{API-Gateway} instead of \textit{Customers}, shows an intriguing outlier. Despite ME-llm2vec correctly matching the "Pet" related classes, it failed with \textit{PetRequest}. its function as a Request object, which is typically associated with the Gateway pattern, is a potential reason. Notably, ME-llm2vec successfully identified \textit{API-Gateway} classes, differentiating them from \textit{Customers}. We find this interesting because \textit{API-Gateway} includes classes representing various bounded contexts, often causing confusion in other models. ME-llm2vec recognized these classes' distinct purpose, grouping them together despite their diverse domains.

% Both \textit{API-Gateway} and \textit{Customers} services contain a "PetType" class. But in \textit{Customers}'s case, this class was closer to the "Specialty" class from \textit{Vets}, which is likely due to nearly identical source code they have. 

\subsection{Discussion}

We designed the analysis component to be as abstract as possible to accommodate the rapidly evolving representation learning landscape. As new and improved embedding models are published, they can be integrated with minimal effort. While our evaluation results show that with ME-LLM2Vec, we can generate highly cohesive and consistent decompositions, one of our objectives is to highlight the potential of Language Models in generating more efficient representations than traditional approaches for the decomposition problem. In fact, MonoEmbed is both a decomposition approach (when considering the full approach) and an embedding model (when using models such as ME-LLM2Vec). These models can be used to enrich existing decomposition approaches. For example, MicroMiner's CodeBERT \cite{trabelsi2023microminer} can be replaced with ME-LLM2Vec and the GNN based methods \cite{desai2021cogcn,yedida2023deeply,mathai2022chgnn,qian2023gdcdvf} can be extended by using ME-LLM2Vec as the encoder. In fact, it can be used as an additional representation type in approaches such as \cite{khaled2022hydecomp,qian2023gdcdvf}. These models can be even extended further by incorporating unstructured inputs (e.g. resources, configurations, documentation) and different PLs.

\subsection{Threats to Validity}
\subsubsection{Internal Validity}
Clustering algorithms and decomposition approaches have hyper-parameters that can affect performance on evaluation benchmarks. To mitigate this threat, we compared their performance with different hyper-parameter inputs across a varied set of evaluation applications.

\subsubsection{External Validity}
To address the threat of our approach to generalize on monolithic applications and PLs, we used a large set of monolithic and microservices applications from related work \cite{kalia2021mono2micro,khaled2022hydecomp,yedida2023deeply,jin2021fosci} to benchmark decomposition approaches. 

\subsubsection{Construct Validity}
This threat can potentially be in the form of the evaluation metrics used in our experiments. In order to mitigate this threat, we employ established metrics in supervised learning tasks (RQ1-3) and different metrics from decomposition research \cite{khaled2022hydecomp,kalia2021mono2micro,jin2021fosci,yedida2023deeply,mathai2022chgnn} (RQ4). 

\section{Conclusion}
We propose, MonoEmbed, a Language Model-based decomposition approach for efficient source code representation in monolithic applications. After evaluating various Pre-Trained Language Models, we fine-tune them for the decomposition task using Contrastive Learning. Our approach significantly improves embedding vector quality, creating similar representations for classes within the same microservices while those in different microservices are distinct. Experiments with clustering algorithms demonstrate that Affinity Propagation with our models better reflects original microservices. Compared to existing benchmarks, MonoEmbed generates more consistent and cohesive decompositions and scales effectively with larger applications. In future work, we would like to experiment with integrating our models with existing decomposition methods and explore new granularities and modalities.

\bibliographystyle{acm}
\bibliography{main}

\end{document}